\documentclass[graybox]{svmult}

\usepackage{mathptmx}       
\usepackage{helvet}         
\usepackage{courier}        
\usepackage{type1cm}        
\usepackage{makeidx}         
\usepackage{graphicx}        
\usepackage{multicol}        
\usepackage[bottom]{footmisc}

\usepackage{cite}

\makeindex             

\begin{document}

\title*{Directional detection of galactic dark matter }
\author{F. Mayet, J. Billard and D. Santos}
\institute{LPSC, Universit\'e Joseph Fourier Grenoble 1,  CNRS/IN2P3, Institut Polytechnique de Grenoble, Grenoble, France, 
 \email{mayet@lpsc.in2p3.fr}}

%
\maketitle

\abstract{Directional detection is a promising Dark Matter search strategy.  
Taking advantage on the rotation of the Solar system around the galactic center through the Dark Matter halo, 
it allows to show a direction dependence of WIMP events that may be a powerful tool to identify genuine WIMP events as such. 
Directional detection strategy requires  the simultaneous measurement of the energy  and the 3D track  of low energy recoils,  which is a common challenge for all current projects of directional detectors. 
}

\section{Introduction}
Since the pionner paper of D.~N.~Spergel~\cite{spergel}, the contribution of 
directional detection to the field of Dark Matter has been adressed through a wealth of 
studies~\cite{albornoz,billard.disco,billard.exclusion,billard.ident,billard.profile,henderson,morgan1,morgan2,
copi1,copi2,copi3,green1,green2,green.disco,Alves:2012ay,Lee:2012pf,Bozorgnia:2011vc,Bozorgnia:2012,Creswick:2010dm,Kuhlen:2012fz,
Lisanti:2009vy,Alenazi:2007sy,Gondolo:2002np}. 
Depending on the unknown WIMP-nucleon cross section, directional detection may be used to : 
exclude Dark Matter \cite{billard.exclusion,henderson}, reject the isotropy 
hypothesis \cite{morgan1,morgan2,copi1,copi2,copi3,green1,green2}, discover galactic Dark Matter with a high 
significance \cite{billard.disco,billard.profile,green.disco} or constrain WIMP and halo 
properties \cite{billard.ident,Alves:2012ay,Lee:2012pf}.

\section{Experimental issues}
Directional detection   requires  the simultaneous measurement of the recoil energy ($E_R$) and the 3D track ($\Omega_R$) of low energy recoils, 
thus  allowing to evaluate the double-differential spectrum $\mathrm{d}^2R/\mathrm{d}E_R\mathrm{d}\Omega_R$ down to the energy threshold.
This can be achieved with low pressure Time Projection Chamber (TPC) detectors  and 
there is a worldwide effort toward the development of a large TPC devoted to directional detection \cite{white}. All current 
projects \cite{dmtpc,drift,d3,mimac,newage} face common challenges amongst which the 3D reconstruction  
of low energy tracks ($\mathcal{O}(10-100)$ keV) is the major one as it includes various experimental issues such as 
sense recognition, angular and energy resolutions and energy threshold. Their effect on the discovery potential of 
forthcoming directional detectors has been fully studied in \cite{green2,billard.profile,billard.exclusion}. The goal of 3D track reconstruction is to 
retrieve, for each track, the initial recoil direction ($\theta, \phi$) and the vertex (X, Y and Z) of the 
elastic scatterring interaction. Difficulties come from the fact that the recoil energy is low (below 100 keV) and the 
track length is small (below 10 mm).\\ 
Recently, a likelihood method dedicated to 3D track reconstruction has been proposed
\cite{billard.track} and applied to the MIMAC detector. The conclusion is
as follows. A good spatial resolution can be achieved, {\it
i.e.}  sub-mm in the anode plane and cm along the drift axis, opening the possibility to perform a fiducialization of 
directional detectors. The angular resolution is shown to range between 20$^\circ$ to 80$^\circ$, depending on the recoil energy, 
which is however enough to achieve a high significance discovery of Dark Matter. On the contrary, the sense recognition capability 
of directional detectors depends strongly on the recoil energy and the drift distance, with small efficiency values (50\%-70\%). 
Moreover, electron/nuclear recoil discrimination may be achieved thanks to a multivariate data analysis based on discriminant 
observables related to the track topology \cite{billard.discri}.

\section{Directional detection : a powerful tool ?}
Taking advantage on the rotation of the Solar system around the galactic center through the Dark Matter halo, 
directional detection strategy enables the use of the expected direction dependence of WIMP events. Indeed,  WIMP event distribution 
should present an excess in the direction of motion of the Solar system, which happens to be roughly in the 
direction of the Cygnus constellation ($\ell_\odot = 90^\circ,  b_\odot =  0^\circ$ in galactic coordinates). 
As the background distribution is expected to be isotropic in the galactic rest frame, one expects a clear and unambiguous 
difference between the WIMP signal and the background one.\\
Beyond the exclusion strategy  \cite{henderson,billard.exclusion}, directional detection may be used to prove that the 
directional data are not compatible with background. With the help of unbinned likelihood method \cite{copi3} or non-parametric 
statistical tests on unbinned data \cite{green2}, it has been shown that a few number of events 
${\cal O} (10)$ is required to reject the isotropy, and hence prove the data are not compatible with the expected background.\\

Directional detection may also be used to discover Dark Matter \cite{billard.disco,billard.profile,green.disco}. In particular, 
the method proposed in \cite{billard.disco} is a blind likelihood analysis, the proof of discovery being the fact that 
the signal points to the direction of the Cygnus constellation. The main direction 
of the incoming events matches the expected direction within 10$^\circ$ to 20$^\circ$ (68\% CL), thus providing an unambiguous signature 
of their origin. Even at low exposure,  a high significance discovery is achievable,  in the presence of 
a sizeable background contamination and for various detector configurations \cite{billard.profile}. 
The goal of this new approach is thus 
not to reject the background hypothesis, but rather to identify a genuine WIMP signal as such. Moreover, one of the strength of 
directional detection strategy is the possibility to go  beyond the standard Dark Matter halo 
paradigm \cite{billard.profile} by accounting for most astrophysical uncertainties (see \cite{Green:2011bv} for a review).\\
Moreover, directional detection provides a powerful tool to explore neutralino Dark Matter models as most MSSM configurations, 
and to a lesser extent for NMSSM ones, with a  neutralino lighter than 
200 $\rm GeV/c^2$ would lead  to a significance greater  than 3$\sigma$ (90\% CL) 
in a 30 kg.year CF$_4$ directional detector \cite{albornoz}.\\

For high WIMP-nucleon  cross section, it is also possible to go further  by constraining WIMP and halo 
properties \cite{billard.ident}. A high dimensional multivariate 
analysis of forthcoming directional data would enable the identification of WIMP Dark Matter. Indeed, a 30 kg.year $\rm CF_4$ 
directional detector would allow us to constrain  the  WIMP properties, both from particle physics (mass and cross section) and 
galactic halo (velocity dispersions). This is a key advantage for directional detection with respect to direction-insensitive strategy. 
Indeed, as the velocity dispersions are set as free parameters, induced bias due to wrong model assumption should be
avoided. This is for instance the effect observed in \cite{green.jcap0708}, with a systematic downward shift of the estimated cross section, when assuming a standard isotropic velocity
distribution fitting model whereas the input model is a triaxial one.\\
Recent studies have also shown the possibility to observe features in the directional signal 
\cite{Bozorgnia:2011vc,Bozorgnia:2012}, such as aberrations and rings, 
which could be used as additional indications in favor of a Dark
Matter discovery.\\

The use of directional detection to constrain the astrophysical properties of Dark Matter has received much interest in the past years.
Recent results of N-body simulation \cite{nbody.VL2,nezri,lisanti1,vogel,Read:2008fh,Read:2009,Purcell:2009yp} seem to favor the presence of 
substructures in the Milky Way halo, such as Dark Matter tidal streams 
(spatially localized), debris flow (spatially homogenized but with velocity substructures) and a 
co-rotating  dark disk. Such components of the local Dark Matter distribution  may  lead to distinctive features 
in the expected directional signal \cite{Lee:2012pf,Kuhlen:2012fz,Green:2010gw}, although the conclusion depends strongly of their unknown
properties. As a matter of fact, constraining their properties remains however a challenging task requiring for 
instance a very low threshold.\\
In a so-called {\it post-discovery era}, 
meaning the WIMP mass is supposed to be known to sufficient precision, it has been shown that directional detection may be used to infer 
astrophysical properties of Dark Matter, namely its phase space distribution in the solar neighborhood \cite{Alves:2012ay}.
In particular, a parametrization of the functional form of the Dark Matter distribution is proposed, 
avoiding to rely on ansatzes. In this case, the coefficients of its decomposition in moments of a model independent basis are 
the measurable quantities in a directional experiment.

\section{Conclusion}
A low exposure $CF_4$ directional detector would offer a unique opportunity in Dark Matter search, by leading, depending on 
the value of the unknown  WIMP-nucleon cross section, either to a conclusive exclusion, a high significance discovery 
of galactic Dark Matter or even  an estimation of the WIMP properties.  
However, several key experimental issues need to be addressed to achieve these physical goals, both on the detector side    
and on the data analysis one.

\end{document}